\documentclass{PoS}
\usepackage{amsfonts,amsmath,amssymb,amsbsy,bm,graphicx,makeidx,multicol,color}

\title{Neutrino electromagnetic properties: a window to new physics - II}

\ShortTitle{Electromagnetic neutrino window to new physics}

\author{\speaker{Alexander Studenikin}$^{ab}$\\
\llap{$^a$}Department of Theoretical Physics, Moscow State University\\
119992 Moscow, Russia\\
\llap{$^b$} Joint Institute for Nuclear Research \\
Dubna 141980, Moscow Region, Russia\\
E-mail: \email{studenik@srd.sinp.msu.ru}}

\abstract{There is merely a short note on the selected issues of neutrino electromagnetic properties with focus on effects of {\it new physics}. The meaning of \ ``{\it new physics}''  is twofold:
1) a massive neutrino have nonzero electromagnetic properties that can be considered as manifestation of {\it new physics} beyond the Standard Model, and 2) in studies   of   neutrinos         electromagnetic interactions  new effects are  predicted that can lead to new phenomena accessible for observations.}

\FullConference{The European Physical Society Conference on High Energy Physics\\
		5-12 July, 2017\\
		Venice}

\begin{document}
\section{Electromagnetic properties of neutrino}
1. After the Nobel Prize of 2015 in physics was awarded to Arthur McDonald and Takaaki Kajita for
the discovery of neutrino oscillations, there can be no doubt that neutrinos are massive particles.
It has been know since quite many years \cite{Fujikawa:1980yx} that massive neutrinos should have nonzero
magnetic moments. Although up to now there are no indications in favour of nonzero neutrino electromagnetic properties, neither from terrestrial experiments nor from astrophysical observations, the electromagnetic properties is one of popular issues related to neutrinos and this
problem has been discussed many times in recent literature (see, for instance,
\cite{Raffelt:2000kp}-\cite{Akhmedov:2014kxa}
%,Nowakowski:2004cv,Wong:2005pa, Balantekin:2006sw,
%Giunti:2008ve,Studenikin:2008bd,Broggini:2012df,}
). A complete review on neutrino electromagnetic properties and neutrino electormagnetic interactions is given in \cite{Giunti:2014ixa}.
In \cite{Studenikin:2016ykv,Studenikin:2016zdx} an addendum to previous reviews are provided and the most recent new aspects and prospects related to neutrino electromagnetic interactions that have appeared after publication of the review paper \cite{Giunti:2014ixa} are discussed.

It has been often claimed (see, for instance, \cite{Studenikin:2008bd}) that neutrino electromagnetic properties open a window to {\it new physics}. In this short note we would like to justify this statement and to focus a discussion on to what extend and what kind of {\it new physics} neutrino electromagnetic properties do communicate with.

\section{Neutrino magnetic moment}

 Consider the magnetic moment as the most well theoretically appreciated and experimentally studied (constrained) electromagnetic characteristic of neutrinos.  Within the initial formulation of the Standard Model neutrinos are massless particles with zero magnetic moment. Thus, the would be nonzero neutrino magnetic moment regardless of its value should indicate the existence of {\it new physics} beyond the Standard Model. Indeed, as it has been shown in \cite{Fujikawa:1980yx} a minimal extension of the Standard Model with right-handed neutrinos yields for the diagonal magnetic  moment of a Dirac neutrino to be proportional to the neutrino mass $m_i$,
\begin{equation}\label{mu_D}
    \mu^{D}_{ii}
  = \frac{3e G_F m_{i}}{8\sqrt {2} \pi ^2}\approx 3.2\times 10^{-19}
  \Big(\frac{m_i}{1 \ \mathrm{eV} }\Big) \mu_{B},
\end{equation}
where $\mu_B$ is the Bohr magneton. For Majorana neutrinos the diagonal
magnetic moments are zero in the neutrino mass basis and only transition (off-diagonal) magnetic
moments  $\mu^{M}_{ij}$ ($i\neq j$) can be nonzero in this case.

The value of neutrino magnetic moment (\ref{mu_D}) is several orders of magnitude smaller than the present
experimental limits if to account for the existed constraints on
neutrino masses. Note that in general transition magnetic moments are even smaller due to the GIM cancelation mechanism.

The best laboratory upper limit on neutrino magnetic moment has been obtained by the GEMMA collaboration that investigates the reactor antineutrino-electron scattering at the Kalinin Nuclear Power Plant (Russia) \cite{GEMMA:2012}. Within the presently reached electron recoil energy threshold of $T \sim 2.8 $ keV the neutrino magnetic moment is bounded from above by the value
\begin{equation}\label{mu_bound}
\mu_{\nu} < 2.9 \times 10^{-11} \mu_{B} \ \ (90\% \ \mathrm{C.L.}).
\end{equation}
This limit, obtained from unobservant distortions in the recoil electron energy
spectra, is valid for both Dirac and Majorana neutrinos and for both diagonal
and transition moments.
The most recent stringent constraint on the electron effective magnetic
moment
\begin{equation}
{\mu}_{\nu_e}\leq 2.8 \times
10^{-11} \mu _B
\end{equation}
has been reported by the Borexino Collaboration \cite{Borexino:2017fbd}.

A strict astrophysical bound on the neutrino magnetic moment is provided by the observed properties of globular cluster stars and amounts to \cite{Raffelt-Clusters:90} (see also \cite{Viaux-clusterM5:2013,Arceo-Diaz-clust-omega:2015})
\begin{equation}\label{mu_bound_astr}
\Big( \sum _{i,j}\left| \mu_{ij}\right| ^2\Big) ^{1/2}\leq (2.2{-}2.6) \times
10^{-11} \mu _B.
\end{equation}
This stringent astrophysical constraint on neutrino magnetic moments is applicable to both Dirac and Majorana neutrinos.

There is a huge gap of many orders of magnitude
between the present experimental limits on
neutrino magnetic moments and the prediction of a minimal
extension of the Standard Model. Therefore, if any direct
experimental confirmation of nonzero neutrino magnetic moment were
obtained in a reasonable future, it would open a window to {\it new physics} beyond a minimal
extension of the Standard Model.

 Much larger values for a neutrino magnetic moments are predicted in
different other extensions of the Standard Model. However, there is a general
problem for a theoretical model of how to get large
magnetic moment for a neutrino and simultaneously to avoid an
unacceptable large contribution to the neutrino mass (see the corresponding discussion in \cite{Giunti:2014ixa} and references therein).
If a contribution to the neutrino magnetic moment of an order $\mu_{\nu}
\sim \frac{eG}{\Lambda}$ is generated by physics beyond a minimal
extension of the Standard Model at an energy scale characterized by
$\Lambda$, then the corresponding contribution to the neutrino mass
is
%\begin{equation}\label{mu_Lambda}
$\delta m_{\nu} \sim \frac{\Lambda ^2}{2m_e}\frac{\mu_{\nu}}{\mu_B}=
\frac{\mu_{\nu}}{10^{-18}\mu_B}\Big(\frac{\Lambda}{1 \ Tev}\Big)^2\
eV$.
%\end{equation}
Therefore, a particular fine tuning is needed to get large value
for a neutrino magnetic moment while keeping the neutrino mass
within experimental bounds. Different possibilities to have large magnetic moment for a
neutrino were considered  in the literature (see in \cite{Giunti:2014ixa}).

A general and termed model-independent upper bound on the Dirac neutrino
magnetic moment, that can be generated by an effective theory beyond
a minimal extension of the Standard Model, has been derived in
\cite{Bell:2005kz}: $\mu_{\nu}\leq
10^{-14}\mu_B$. Note that the corresponding limit for transition moments of Majorana neutrinos is much weaker \cite{Bell:2006wi}. Thus, the value of a neutrino magnetic moment once observed
experimentally at the level not less than $\mu_{\nu}\sim10^{-14}\mu_B$ would provide
information on the nature of neutrinos. This can be also considered as a view
on the realm of {\it new physics}.

\section{Neutrino electric moment}

  From the most general form of the neutrino
electromagnetic vertex function $\Lambda _{ij}(q^{2})$
(see for detailed discussion \cite{Giunti:2014ixa})
there are  three other sets (in addition to the magnetic moments $\mu_{ij}$) of electromagnetic
characteristics that determine a neutrino coupling with real photons ($q^{2}=0$). They are namely the dipole electric moments $\epsilon_{ij}$, anapole moments $a_{ij}$
and millicharges $q_{ij}$. In the theoretical framework with $CP$ violation a neutrino
can have nonzero electric moments $\epsilon_{ij}$. In the laboratory neutrino
scattering experiments for searching the neutrino magnetic moment (like, for instance,
the mentioned above GEMMA experiment) the electric moment contributions interfere with
those due to magnetic moments. Thus, these kind of experiments also provide constraints
on $\epsilon_{ij}$. The astrophysical bounds (\ref{mu_bound_astr})
%Viaux-clusterM5:2013,Arceo-Diaz-clust-omega:2015}
are also applicable for constraining $\epsilon_{ij}$ \cite{Raffelt-Clusters:90}-
\cite{Arceo-Diaz-clust-omega:2015}.

\section{Neutrino electric millicharge}

There are extensions of the Standard Model that allow for nonzero
neutrino electric millicharges. This option can be provided by
not excluded experimentally possibilities for hypercharhge dequantization or
another {\it new physics} related with an additional $U(1)$ symmetry
peculiar for extended theoretical frameworks. Neutrino millicharges
are strongly constrained on the level $q_{\nu}\sim 10^{-21} e_0$
($e_0$ is the value of an electron charge) from neutrality of the hydrogen atom.

 A nonzero neutrino millicharge $q_{\nu}$ would contribute to the neutrino electron scattering in the terrestrial experiments. Therefore, it is possible to get bounds on $q_{\nu}$ in the reactor antineutrino experiments GEMMA. The most stringent constraint using the GEMMA data  is $q_{\nu}\leq 1.5 \times 10^{-11} e_0$ \cite{Studenikin:2013my} (see also \cite{Patrignani:2016xqp}).

A neutrino millicharge might have specific phenomenological consequences
in astrophysics because of new electromagnetic processes are opened
due to a nonzero charge. Following this line, the most stringent astrophysical constraint on neutrino millicharges
$q_{\nu}\leq 1.3 \times 10^{-19} e_0$ was obtained in \cite{Studenikin:2012vi}. This bound
follows from the impact of the neutrino star turning mechanism ($ST\nu$) \cite{Studenikin:2012vi} that can be charged as a {\it new physics} phenomenon end up with a pulsar rotation frequency shift engendered by the motion of escaping from the
star neutrinos on curved trajectories due to millicharge interaction with a constant
magnetic field.

\section{Neutrino charge radius}

Even if a neutrino millicharge is vanishing, the electric form factor
$f_{ij}^{q}(q^{2})$ can still contain nontrivial information about
neutrino electromagnetic properties. The corresponding electromagnetic characteristics is
determined by the derivative of $f_{ij}^{q}(q^{2})$ over $q^{2}$  at
$q^{2}=0$ and is termed neutrino charge radius,
%\begin{equation}
%\label{nu_cha_rad_1}
${r}_{ij}^{2}
=-
6
%\left.
\frac{df_{ij}^{q}(q^{2})}{dq^{2}} \
_{\mid _ {q^{2}=0}}
$.
%\end{equation}
A neutrino charge radius (that is indeed the charges radius squared) contributes to the neutrino scattering cross section on electrons and thus
can be constrained by the corresponding laboratory experiments \cite{Bernabeu:2004jr}.
In all (see, for instance, \cite{Giunti:2014ixa}) but one previous studies it was claimed
 that the effect of the neutrino
charge radius can be included just as a shift of the vector coupling constant $g_V$
in the weak
contribution to the cross section.
However, as it has been recently illustrated, in \cite{Kouzakov:2017hbc} within the direct calculations of
the cross section accounting for all possible neutrino electromagnetic characteristics
and neutrino mixing, this is not the fact. The neutrino charge radius dependence of the cross section
indeed is more complicated and there is, in particular, the dependence on the interference terms of the type
$g_{V}{r}_{ij}^{2}$ that can't be obtained just only by the corresponding shift of the constant $g_V$.

\section{Neutrino spin precession in magnetic field}

One of an important phenomenon among several processes of neutrino electromagnetic interacts is neutrino spin and spin-flavour precession in magnetic fields. The origin of these effects is the neutrino magnetic moment interaction with a transversal magnetic field determined by $\mu_{\nu}B_{\perp}$. The neutrino spin precession
in a transverse magnetic field can result in the neutrino helicity flip that can have important phenomenological consequences because an active neutrino $\nu_{L}$ can be converted to a sterile one $\nu_{R}$ in invironments with a magnetic field.  The precession $\nu_{eL}\rightarrow \nu_{eR}$ in the magnetic field of the Sun was first considered in \cite{Cisneros:1970nq}, a similar effect in magnetic fields of supernovae and neutron stars came into sight for the first time  in \cite{Fujikawa:1980yx}.

\section{Neutrino spin precession in transversal matter currents}

There is a phenomenon of {\it new physics} related to the neutrino spin precession in magnetic fields. For many years, until  2004, it was believed that a neutrino helicity precession
and the corresponding spin oscillations can be induced by the neutrino magnetic
interactions with the transversal magnetic field. A new and very interesting possibility for neutrino spin
(and spin-flavour) oscillations engendered by the neutrino interaction with matter background was
proposed and investigated in \cite{Studenikin:2004bu}. It
was shown that neutrino spin oscillations
can be induced not only by the neutrino interaction with a  magnetic field but also by neutrino interactions with matter in the case when there is a transversal matter current (or a transversal matter matter polarization). The is no need for neutrino magnetic moment interaction in this case. The origin of the oscillations $\nu_{L}\Leftrightarrow\nu_{R}$ in the transversal matter currents $j_{\perp}$ is the neutrino weak interactions with moving matter and the corresponding mixing between neutrino states $\nu_{L}$ and $\nu_{R}$  is determined by $G_F j_{\perp}$. This new effect has been explicitly highlighted in \cite{Studenikin:2004bu,Studenikin:2004tv}, recently the existence of this effect was confirmed in \cite{Kartavtsev:2015eva}. For historical notes reviewing studies and the detailed derivation of the discussed effect see \cite{Studenikin:2016ykv,Studenikin:2016zdx} and \cite{Studenikin:2016iwq}.

\section{Conclusions and future prospects}

 The foreseen progress in constraining neutrino electromagnetic characteristics is related, first of all, with the expected new results from the GEMMA experiment measurements of the reactor antineutrino cross section on electrons at Kalinin Power Plant. The new set of data is expected to arrive next year. The electron energy threshold will be as low as $350 \ eV$ ( or even lower, $\sim 200 \ eV$). This will provide possibility to test the neutrino magnetic moment on the level of $\mu_\nu \sim 0.9 \times 10^{-12} \mu_B$ and also to test the millicharge on the level of $q_{\nu} \sim 1.8 \times 10^{-13} e_0$ \cite{Studenikin:2013my}. For the next future, presently it seems unclear whether further progress in constraining the neutrino electromagnetic characteristics would be achievable with this type of the reactor antineutrino  experiment. In this concern, a rather promising claim was made in  \cite{deGouvea:2012hg,deGouvea:2013zp}. It was shown that even much smaller values of the Majorana neutrino transition moments
would probably be tested in future high-precision experiments with the astrophysical neutrinos.  In particular,
observations of supernova fluxes  in the JUNO  experiment (see \cite{An:2015jdp}-
%,Giunti:2015gga,
\cite{Lu:2016ipr})  may reveal the effect of  collective  spin-flavour oscillations  due to the Majorana neutrino transition moment $\mu^{M}_\nu \sim 10^{-21}\mu_B$.

To conclude, the existing current constraints on the flavour neutrino charge radius ${r}_{e,\mu,\tau}^{2}\leq 10^{-32} - 10^{-31} \ cm ^2$  from the scattering experiments differ only by 1 to 2
orders of magnitude from the values ${r}_{e,\mu,\tau}^{2}\leq 10^{-33} \ cm ^2$ calculated within the minimally extended Standard Model with right-handed neutrinos \cite{Bernabeu:2004jr}. This indicates that the minimally extended Standard Model neutrino charge radii could be experimentally tested in the near future. Note that there is a need to re-estimate experimental constraints on ${r}_{e,\mu,\tau}^{2}$  from the scattering experiments following new derivation of the cross section \cite{Kouzakov:2017hbc} that properly accounts for the interference of the weak and charge radius electromagnetic interactions and also for
the neutrino mixing.

\section{Acknowledgements}
This work was supported by the Russian Foundation for Basic Research under grants
No.~16-02-01023\,A and No.~17-52-53133\,GFEN\_a.

\end{document}